# Improving the Quality of FORS2 Reduced Spectra

Sabine Moehler[1]
C. E. García Dabo[1]
Henri Boffin[1]
Gero Rupprecht[1]
Ivo Saviane[1]
Wolfram Freudling[1]

[1] ESO

The FORS2 instrument is one of the most widely used and productive instruments on the Very Large Telescope. This article reports on a project to improve the quality of the reduced FORS2 spectra that can be produced with the software provided by ESO. The result of this effort is that spectra of significantly higher quality can now be produced with substantially lower effort by the science user of the data.

## Introduction

The two FOcal Reducer/low dispersion Spectrograph (FORS) instruments (Appenzeller et al., 1998; Rupprecht et al., 2010) are among the most popular instruments at the VLT. They are multi-mode instruments covering the optical wavelength range from 330 to 1100 nm, which have been in operation since 1 April 1999 (FORS1, decommissioned on 31 March 2009) and 1 April 2000 (FORS2), respectively. With their many modes, the FORS instruments are real Swiss Army knives. They support direct-imaging modes with narrow- and broadband filters, long-slit and multi-object spectroscopy, polarimetry and high time-resolution observations.

It is therefore not surprising that up until now, more than 1880 refereed papers have been published, based on data from FORS1 and FORS2. Among the scientific highlights discovered with the FORS instruments are conclusive evidence for the link between gamma-ray bursts and hypernovae (Hjorth et al., 2003) and the first ground-based transmission spectrum of a super-Earth planet (Bean et al., 2010). They have also enabled the discovery of the furthest quasar yet found, at redshift 7.1 (Mortlock et al., 2011) and the furthest gamma-ray burst at redshift 8.2 (Tanvir et al., 2009).

FORS2 observes objects closer to us as well, for example, allowing astronomers to find close binaries inside some magnificent planetary nebulae (Boffin et al., 2012), or permitting detailed studies of comets and asteroids, and even "rediscovered" life on Earth (Sterzik et al., 2012). For all these reasons FORS2 is still among the two most highly demanded VLT instruments. It is also behind some iconic astronomical pictures.

The FORS instruments produce a wide variety of complex data products that reflect their many different modes and options. Reducing these data, in particular spectroscopic data, is often quite a challenge. Since late 2006 ESO has provided pipeline software that can be used to process science data. Unfortunately, the products from the unsupervised reduction with that software did not always produce satisfactory results, and a variety of external software tools were necessary to produce science-ready results. In addition, using this pipeline required a significant effort of data organisation by the user. To remedy this situation, we embarked on a two-year effort to improve the FORS2 spectroscopic data products, looking at all aspects of the instrument operation that have an effect on the final data quality. In particular, we verified the current calibration plan, upgraded many aspects of the ESO pipeline, and produced a workflow in the ESO Reflex environment[1] that allows both "black box" and interactive execution of the FORS pipeline. The result of this effort is a major improvement in the quality of the spectra as they come out of the FORS pipeline. In this article, we summarise the most important aspects of these upgrades. The improved pipeline can be downloaded[2].

## Spectroscopy with FORS2

FORS2 spectroscopy consists of three modes: classical long-slit spectroscopy with slits of 6.8 arcminutes length and predefined widths between 0.3 and 2.5 arcseconds (long-slit spectroscopy mode, [LSS]); multi-object spectroscopy with 19 slitlets of 20–22 arcseconds length each and arbitrary width created by movable slit blades (MOS mode); and multi-object spectroscopy using masks with slitlets of almost arbitrary length, width, shape and angle (MXU mode). There are 15 grisms with resolutions (for a 1-arcsecond slit) from 260 to 2600, which may be combined with three different order-separating filters to avoid second-order contamination. More information about the instrument is available on the FORS web page[3].

## Reflex workflow

Like all other ESO pipelines, the FORS pipeline can be executed with the command line interface "esorex". However, the organisation of FORS2 data for the various data reduction steps and the correct transfer of products from one recipe to the next can be cumbersome and error-prone. Automatic pipeline processing of FORS2 data will, in general, provide satisfactory results for a large fraction of all data. However, in many cases fine tuning of some parameters of the pipeline will further improve the results, and in some cases, such fine tuning is essential. For example, multi-object spectroscopy can sample very different parts of a dispersion relation, so that the default parameters for the dispersion relation may not be best suited for a specific observation.

In order to permit efficient processing of FORS2 data, including the interactive modification of pipeline parameters, we created a Reflex workflow (Freudling et al., 2013), which is shown in the screenshot in Figure 1. With this workflow, it is possible to reduce FORS2 spectroscopic data with the single push of a button. It also allows intermediate results to be inspected, parameters modified, and the impact of such modifications on the results to be assessed immediately.

This workflow organises the data according to configurable rules, allows the user to select the data to be processed and then performs the processing in the correct sequence, including the handover of products from one recipe to the next. Interactive windows like the one shown in Figure 2 allow the results to be checked and recipe parameters fine-tuned if necessary. It is also possible to switch off all interactivity and process large amounts of data in an automatic and unsupervised way.

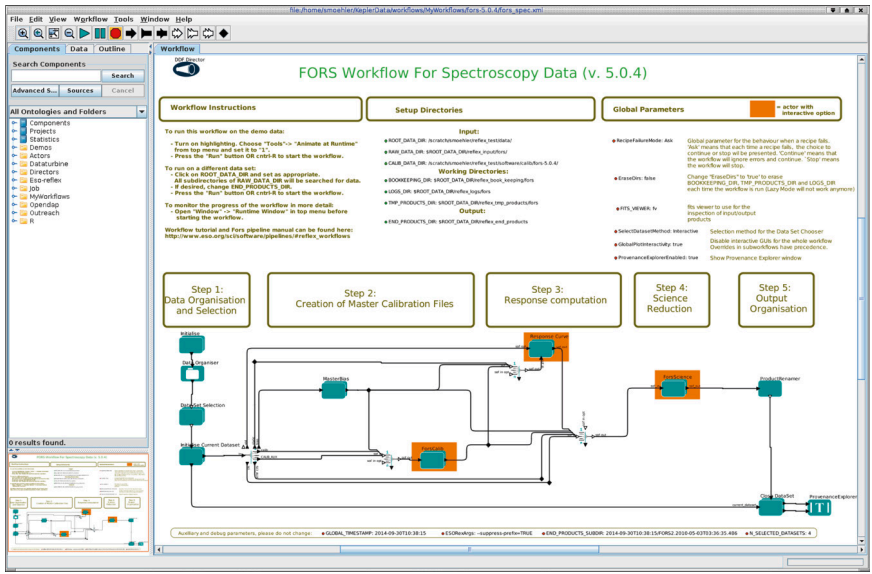

Figure 1. (Above) Screenshot of the FORS2 spectroscopy workflow. The upper part contains instructions and configuration settings, while the lower part contains the actual workflow.

Figure 2. (Below) Screenshot of one of the interactive windows of the FORS2 spectroscopy workflow. The windows provide information about the quality of the wavelength calibration, slit tracing and flat-field normalisation, while the tabs on the right allow the user to change parameters and rerun the recipe.

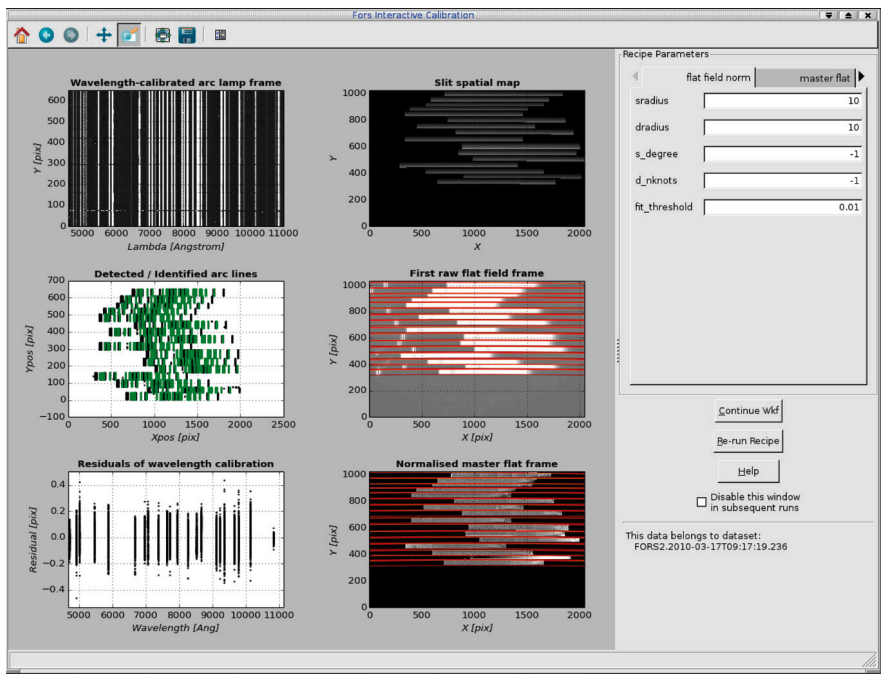

### Error propagation

The full reduction chain has been retrofitted with estimates of the uncertainties based on error propagation starting from the raw data. The initial uncertainties of all raw images are based on the photon shot noise and the detector noise model. Subsequent propagation includes the uncertainties of the biases, flats and spectra extraction.

It should be noted that the systematic errors are not added to the statistical uncertainties. The pipeline gives some idea of the systematic errors as quality control parameters, and they can be used for reference and comparison with the statistical errors.

### Stability of biases and their noise

In the past the FORS2 detectors have sometimes shown significant variations of the bias level on periods of days. To safeguard against the negative effects of such changes, we implemented pre-/over-scan correction, using the median of the pre-scan region to scale the bias level. As FORS2 has only very narrow pre-/over-scan regions (five pixels in the default read-out modes) and some of those tend to be corrupt, we decided to use the median instead of more elaborate descriptions. Due to the small number of useful pre-/over-scan pixels, the read-out noise per frame cannot be determined from these regions. Instead the read-out noise is now determined from the full-bias frames and stored in the master bias. Assuming that it does not change between the bias and the other frames, this value is used as read-out noise for the error propagation along the full reduction chain. To take into account small slopes in the bias level we subtract the full master bias and not just the median of the pre-scan. To reduce the noise of the master bias, the number of bias frames taken during the day was increased from five per read-out mode to twenty. The errors of the master bias are stored in an error extension.

### Wavelength and distortion calibration

The FORS2 spectroscopic pipeline processes the associated flat-field and arc-lamp frames simultaneously in order to provide a consistent description of the distortions along the spatial and dispersion axes respectively. Line identification is performed by pattern matching, and the dispersion relation is fitted by a polynomial, either row by row or using additional information along the spatial axis. We found that the algorithm is generally robust and gives good results for the dispersion relations with polynomial fits of degrees 3 to 5 (depending on the grism), if appropriate parameters and line lists are used. Therefore we revised these ingredients and could thus reduce the random error of the wavelength calibration by about 25%, or from 0.15 to 0.12 pixels on average.

A good quality and homogeneous wavelength calibration is especially important for multi-object observations, where the wavelength ranges covered may differ for different slit (and object) positions, but the results should be comparable across the field of view. The current arc lamps provide few lines at the edges of the FORS2 wavelength range, i.e., below 4000 Å (four lines only) and above 10 000 Å (two lines). In order to improve the situation for future data, we are searching for arc lamps that might provide more lines in these ranges (on the order of ten), while having sufficiently isolated lines to be suitable for low-resolution spectroscopy.

### Flat fielding

Spectroscopic FORS2 flat fields use blue and red flat-field lamps to cover the full spectral range. The flux level and spectral energy distribution (SED) of the blue flat-field lamps can vary by 10% on time scales of minutes, i.e., within one set of flat fields for a given setup, and of course also between such setups. These variations prevent meaningful flat-field stacking by median or sigma-clipping methods, unless the SED (and the flux level) are (at least roughly) normalised before the stacking. Such a normalisation has now been implemented and the average level and SED are put back into the stacked flat, which can then be normalised by smoothing or fitting along the dispersion and/or the spatial axes. The spectrum used to normalise the master flat field along the dispersion axis is retained and used for the flux calibration of volume-phase holographic (VPH) grisms (see below). The errors of the master flat field are derived from the raw flat fields and the master bias and stored in an extension.

Owing to the imperfect slit shapes, the illumination along the slit may vary, which can cause problems for sky correction and the analysis of extended objects. Such variations can amount to about 10% of the flux along the slit for long slits (see Figure 3, red lines). Some principal investigators (PIs) have therefore taken twilight flat fields to correct their observations and we have verified how the illumination gradients in these data compare to screen flat fields.

For that purpose the rectified flat fields (both twilight and screen) were averaged along the dispersion axis and the resulting profiles were compared to each other and to the slit illumination profile of the science data (the latter can be difficult to obtain for bright targets and/or extended objects). We found that the spatial illumination of LSS data is very similar for both kinds of flat fields and the science data. Keeping the spatial gradient of long-slit spectroscopy flat fields (i.e., normalising only along the dispersion axis) will therefore allow the user to correct for the illumination gradient (see Figure 3, black lines).

### Flux calibration

The spectrophotometric calibration of FORS2 data mostly uses flux standard star information from Hamuy et al. (1994). These were derived from low resolution ground-based spectra and binned to 50 Å. The fact that they were observed from the ground means that they contain telluric absorption features, whose strength may not be the same as in the observed FORS2 spectra of these standard stars. In addition, the FORS2 grisms provide a wide range of resolutions, so that, even after binning to 50 Å, the binned flux values for strong absorption lines (like those for white dwarfs) may differ from the tabulated values. We therefore calculated the ratios of observed vs. tabulated flux standard star spectra for all FORS2 flux standard stars (also those provided by other authors) and determined masking regions for stellar features (per star) and telluric features (per grism). These regions are now masked by default when fitting the response curve, but the user can change the masking via recipe parameters and also mask individual points. During this exercise we found that the Oke (1990) flux standard star tables had some problems (as already mentioned by their author) and we decided to remove them from the FORS2 calibration plan.

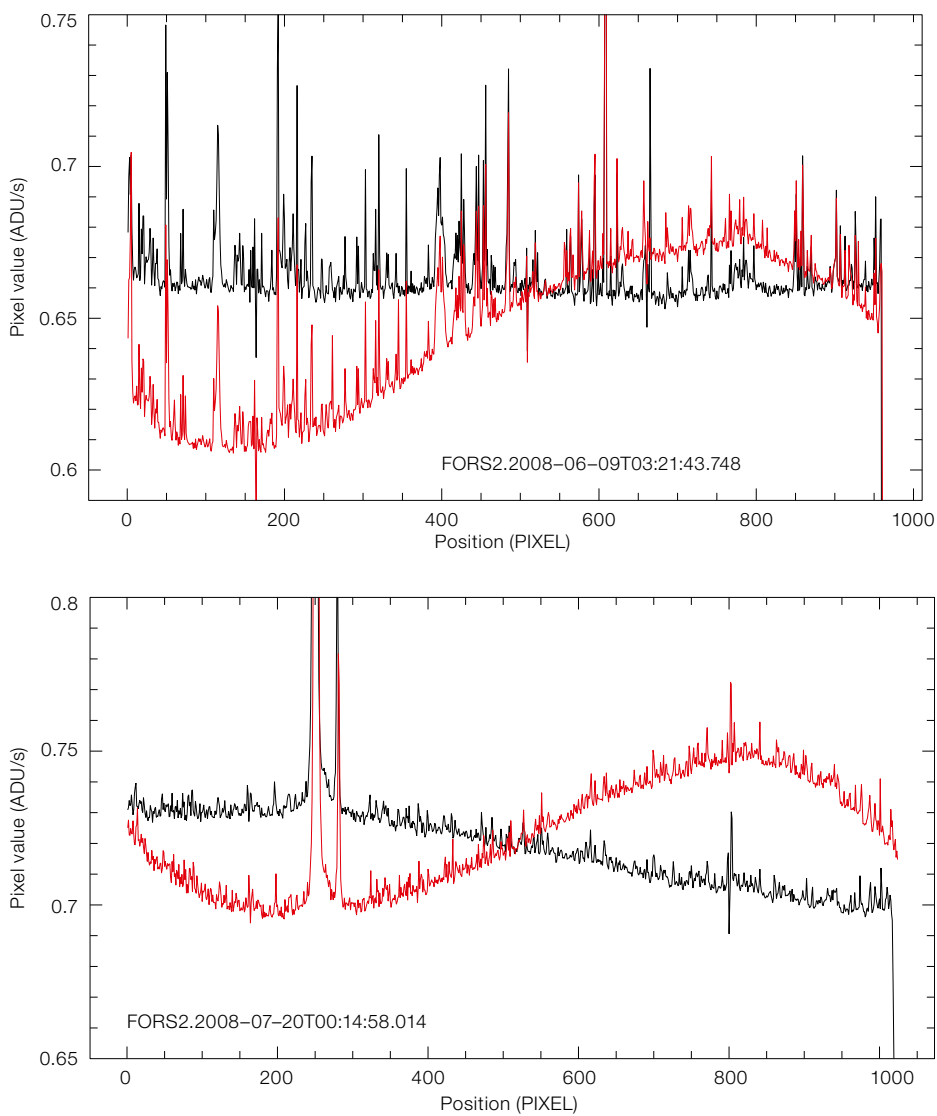

Figure 3. The plots show two examples of the flux variation along the slit as derived by averaging the rectified long-slit spectroscopy science frame along the dispersion axis. The black profile shows the result by using flat fields which retain their spatial gradient, while the red one shows the results for which the spatial profile was eliminated during flat field normalisation.

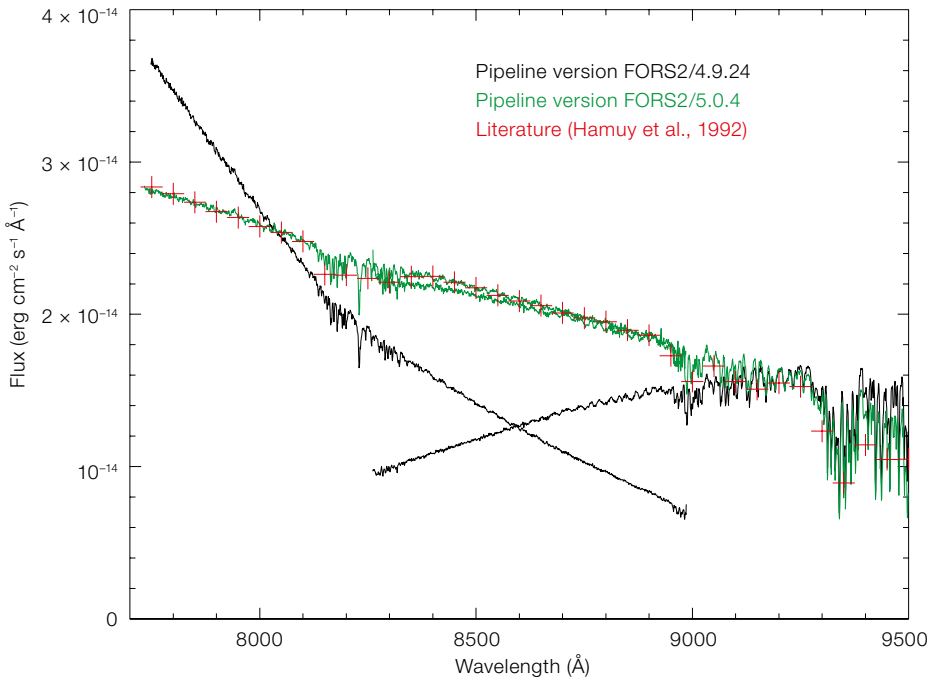

Figure 4. Plot showing the improvement in the flux calibration of data taken with VPH grisms. The spectrum of the standard star LTT3218 was observed with GRIS_1028z+OG590 at ±150 arcseconds from the centre of the field of view, but calibrated with a response derived from observations taken at the centre of the field of view (as done with multi-object spectroscopy science data). The black spectra show the results without flat-field spectrum correction, where the slopes of the calibrated spectra differ for different positions. The green spectra were obtained with the newly implemented flat-field spectrum correction and show the correct slope irrespective of position. The red crosses mark the reference data from Hamuy et al. (1992).

For the volume-phase holographic grisms, which are used for about 50% of the FORS2 spectroscopic data, the response varies with the position on the detector along the dispersion axis. Therefore the flux standard star should be taken with a wide slit at the position of the science target. While this is easily done as part of the calibration plan for LSS data, this requirement poses an obvious problem for multi-object spectroscopy data. For such data the flux-standard star is observed at the centre of the field of view. If the position dependency of the response is not corrected, the flux-calibrated spectra will show systematically distorted slopes (see black lines in Figure 4).

We have fixed this problem by taking into account the response variation with position as seen in the flat-field spectra (Halliday et al., 2004; Milvang-Jensen et al., 2008) and applying it to both the flux-standard star and the science spectra. Then science spectra taken at a large distance from the centre of the field of view can be flux-calibrated using a standard star observed in the centre only (see green lines in Figure 4). This correction relies on the stability of the flat-field lamp in both flux level and spectral energy distribution. The instability of the blue flat-field lamp mentioned above therefore limits the accuracy of this correction at wavelengths below about 4800 Å, i.e., for grisms GRIS_1200B and the blue part of GRIS_1400V. One possible way of safeguarding multi-object spectroscopy taken with these grisms against lamp variations would be to place one slitlet at the centre of the field of view along the dispersion axis in order to have a reference flat-field spectrum to compare to the one used for the standard star.

### Science processing

Given that long-slit spectroscopy is the mode of choice when observing extended objects, it was therefore rather annoying that the pipeline could not correct the spatial distortion along the slit. In order to enable this correction we took arc-lamp and flat-field frames for all available grisms using a dedicated MXU mask and derived global distortion tables, that can now be applied to LSS data to correct the distortion. One example is shown in Figure 5, where the effect of the distortion can be clearly seen at the top of the plot.

If the spatial illumination gradient is kept in the long-slit flat fields and thus corrected in the long-slit science spectra, it was found that the global sky subtraction (which utilises the oversampling of the sky spectrum due to the curvature of the sky lines) works better than the local one, which attempts to fit the spatial flux variations of the sky lines.

### Implementation

The C/C++ MOSCA (MOS Common Algorithms) library has been used to implement most of the algorithms described here. It has been developed at ESO as a common library for MOS instruments (FORS and VIMOS for the time being) and is based on the concept of pattern matching. MOSCA is in itself based on two other ESO projects: CPL (Common Pipeline Library) and HDRL (High-level Data Reduction Library).

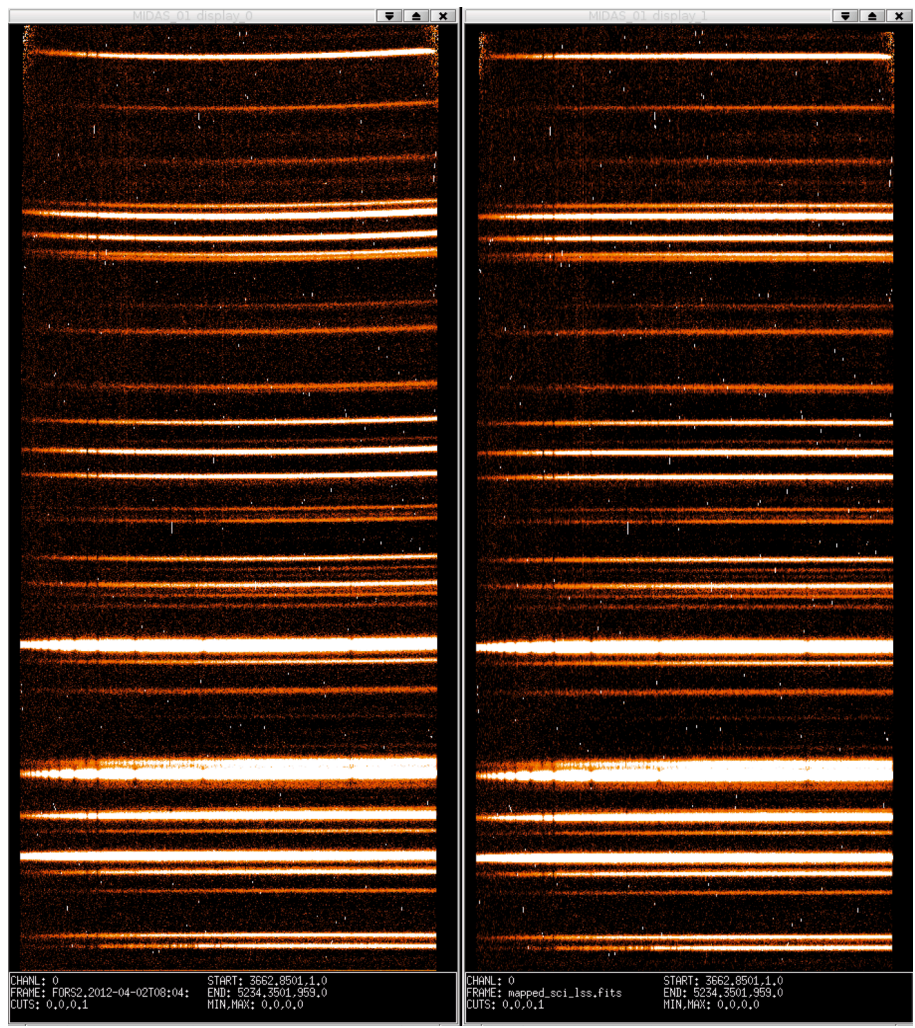

Figure 5. Wavelength calibrated long-slit spectroscopy science frame without (left) and with (right) distortion correction. The spectra are compressed along the dispersion axis to better illustrate the distortion. The position of the centre of the field of view along the y-axis is situated about 10% from the bottom of the frame.

## Verification

In order to verify that our changes in parameters and reference calibration do not cause any deterioration in the quality of the automatic processing, we processed some 400 science datasets observed since 2008 and compared them to the results of the automatic processing by the Quality Control (QC) group, both for science and calibration products. We found that the accuracy of the wavelength calibration had improved by about 25% as mentioned above and the accuracy of the flux calibration had improved by up to 50% for VPH grisms. For the standard (non–VPH) grisms, the improvements were less striking, but there are some grisms here which also profit from more appropriate fit parameters (e.g., GRIS_600B). The signal-to-noise ratio could not be significantly improved as the FORS2 pipeline already uses the optimal extraction method (from Horne, 1986), which is well suited to these kind of data. We found no evidence for any systematic wavelength shift from sky lines above 0.05 pixels.

## Conclusions

The improvements and new features described here will allow users to realise the full potential of their FORS2 spectra more easily. This project illustrates that the quest for improvements in data quality requires not just well-suited pipeline algorithms, but also appropriate calibration plans and reference data. One should therefore always try to look at the complete picture of data and software. The project also showed the inherent problems of working on an instrument with a long history — changes to the calibration plan will always be useful only for future data and will usually not allow existing problems to be fixed.

## Acknowledgements

This research has made use of the NASA Astrophysics Data System Bibliographic Services, the National Institute of Standards and Technology database of atomic lines and the SIMBAD database, operated at CDS, Strasbourg, France. Thanks go to Joe Hennawi, Neil Crighton and Bo Milvang-Jensen for their assistance with the analysis of FORS2 twilight flats and to Alessandro Nastasi for his help with F-VIPGI. We are grateful to Mark Downing for his help with understanding the FORS2 detectors, to Hans Dekker for his explanations of the behaviour of VPH grisms and to Paul Bristow for his support of the physical model test. Special appreciation goes to the late Carlo Izzo, who wrote the original FORS pipeline.

## Links

[1] ESO Reflex: http://www.eso.org/sci/software/reflex/
[2] Download of FORS2 pipeline: http://www.eso.org/sci/software/pipelines/
[3] FORS web page: http://www.eso.org/sci/facilities/paranal/instruments/fors.html